\documentclass[11pt]{article}
\usepackage{moriond,epsfig}

\def\Journal#1#2#3#4{{#1} {\bf #2}, #3 (#4)}

\def\NIMA{{\em Nucl. Instrum. Methods} A}

\def\PRL{\em Phys. Rev. Lett.}

\def\be{\begin{equation}}
\def\ee{\end{equation}}
\def\bea{\begin{eqnarray}}
\def\eea{\end{eqnarray}}

\begin{document}

\vspace*{4cm}
\title{First physics results from the HARP experiment at CERN}

\author{ A. Cervera Villanueva}

\address{University of Geneva, Switzerland}

\maketitle\abstracts{
The first physics results of the HARP experiment are presented. 
We emphasize the high 
performance of the forward part of the apparatus. 
The differential raw pion yield and its efficiency correction up to
polar angles of $250~mrad$ are shown. The analysed setting is 
$12.9~GeV/c$ incident protons in a $5\%$ interacion legth aluminium
target. 
}

\section{Introduction}

The HARP experiment \cite{harp} was designed to perform a systematic and
precise study of hadron production for beam momenta between 1.5 and
$15~GeV/c$ , for target nuclei ranging from hydrogen to lead. 
The detector was located at CERN, in the PS beam, and took 420 million events  
during the years 2001 and 2002.

The physics goals of HARP are to make a measurement of the pion yield
that will enable a quantitative design of the proton driver of a
neutrino factory, 
and to improve the precision of atmospheric neutrino flux
calculations. 
In addition, the energy-range is suitable to measure particle
yields for the prediction of neutrino fluxes for the MiniBooNE
\cite{miniboone} and K2K \cite{k2k} experiments ($8.9$ and $12.9~GeV/c$ respectively). 
To this end a collaboration was set up with these groups. Beam energy settings
and dedicated targets were used to provide the most relevant
measurements.

\begin{figure}[htbp]
\begin{center}
\hspace{0mm} \epsfig{file=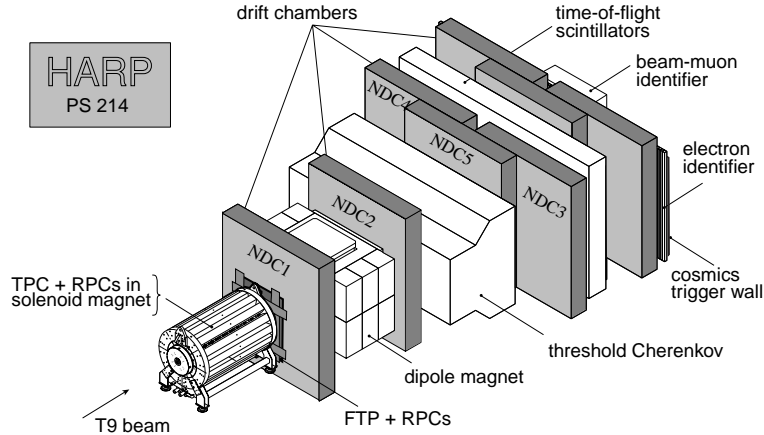,width=10cm}
\end{center}
\caption{Schematic layout of the HARP spectrometer.}
\label{fig:harp}
\end{figure}

A schematic layout of the apparatus is shown in Fig. \ref{fig:harp}. It is
a large acceptance spectrometer, with two distinct regions. 
A forward region (up to polar angles of about $250~mrad$),
where the main tracking devices are a set of drift chambers, 
and where particle identification is possible thanks to 
the combination of a threshold cerenkov, a time-of-flight
wall and an electron identifier. 
In the large-angle region the main tracking and
particle-id detector is a TPC, which is complemented by a set of RPC
detectors for time of flight measurements. 
In addition, a complex beam instrumentation was set up in order to 
establish the nature of the incoming particle, given the rather impure 
beam. The main subsystems are three small timing detectors and two cerenkovs. 
The beam time detectors are also used to estimate the interaction time at the 
target, located inside the TPC. 

Given the immediate interest of the  MiniBooNE and K2K experiments 
in our results, the HARP Collaboration has decided to start the data 
analysis with these two attractive items. In particular, we present in 
this article a first analysis of the data dedicated to the K2K
measurement, using for 
this purpose the forward region of the detector, which is able to cover 
by itself the K2K requirements. 

The organisation is as follows. Section \ref{sec:motivation} introduces the motivation 
of this analysis.  In section \ref{sec:tracking} 
the tracking capabilities of the forward spectrometer are presented. 
Section \ref{sec:pid} is devoted to the particle identification subsystems 
and their performance. Finally, the data analysis is presented 
in section \ref{sec:analysis} and the conclusions in section \ref{sec:conclu}.

\section{Motivation of this analysis} \label{sec:motivation}

One of the main systematic errors on the neutrino oscillation parameters measured by the K2K 
experiment comes from the uncertainty on the far/near neutrino flux ratio. 
This ratio depends on the differential pion production cross section, which is  
essentially measured by a threshold cerenkov (pion monitor) above $E_{\nu}=1~GeV$, but must rely on  
Monte Carlo simulations below that energy. Unfortunately, the oscillation peak is located 
at $E_{\nu} \sim 0.55~GeV$. HARP could precisely estimate the pion differential cross 
section at these energies and subsequently reduce that systematic error.

\begin{figure}[htbp]
\begin{center}
\epsfig{figure=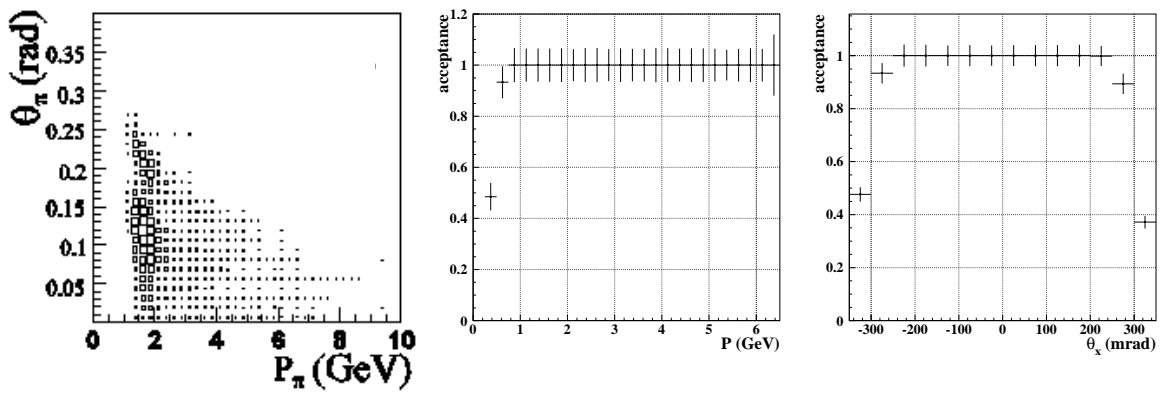,width=16cm} \vspace*{-0.5cm}
\caption{On the left, $(p,\theta)$ distribution for pions producing neutrinos in the 
bin containing the oscillation peak ($0.5-0.75~GeV$). This figure is a
courtesy of the K2K Collaboration. It has been obtained with a full
simulation of the K2K pion production and decay chains. Center and
right,  geometrical acceptance of HARP's dipole magnet for particles arriving to 
the first NDC module (NDC1). On the center, $P$ dependence for
particles with $|\theta_x|<200~mrad$. On the right, $\theta_x$ dependence for
particles with $P>1~GeV/c$. 
\label{fig:k2k_and_acc}}
\end{center}
\end{figure}

Fig. \ref{fig:k2k_and_acc}-left shows the $(p,\theta)$ distribution for
pions producing neutrinos in the 
bin containing the oscillation peak ($0.5-0.75~GeV$). The relevant phase space is 
$1$$<$$p$$<$$8$ $GeV$ and $\theta$$<$$250~mrad$, which is fully covered by HARP's 
forward region as shown in Fig. \ref{fig:k2k_and_acc}.

3.5 million useful events were taken by HARP with incident
$12.9~GeV/c$ protons and an exact replica of the
K2K target, which is a 
$80\times 3~cm$ aluminum tube (2 interaction lengths). Another 6M events were collected 
with a $5\% \lambda$ aluminium target (``K2K thin target'') and a similar
beam in order to decouple reinteraction 
and absorption effects from pure production. The data analysis presented in this article 
is based on this particular setting.

\section{Forward Tracking} \label{sec:tracking}

Tracking of forward going particles is done by a set of drift chambers (NDC)
placed upstream and downstream of the dipole magnet (DIP). The NDC-DIP combination 
allows the momentum measurement by matching track segments 
located at either side of the magnet. 

The chambers were recuperated from the NOMAD
experiment and their properties have been described
elsewhere~\cite{NOMAD_NIM_DC}. 
Each NDC module contains 4 chambers, 
and each chamber 3 planes of wires with 
tilted angles $-5^o$, $0^o$ and $5^o$. The single wire efficiency  is
of the order of $80\%$, and the spatial resolution approximately $340 \ \mu m$.
\footnote{The chambers had a superior performance in NOMAD ($95\%$
  hit efficiency and a resolution a factor of two better). This is
  mainly due to the use of different non-flammable gas mixture
  ($Ar(90\%)-CO_2(9\%)-CH_4(1\%)$), and volatage
  settings ( sense wires held at +1300V and
  potential wires at -2900V). However, one should stress that this
  spatial resolution is more than sufficient 
  for HARP physics}

The reconstruction algorithm builds $2D$ and $3D$  track segments in each NDC module 
(12 hits maximum), which are fitted to a straight line model via a
Kalman Filter fit \cite{Kalman}. 
Afterwards, all the possible matching combinations 
of tracking objects (including unused hits) belonging to different modules 
are performed in order to obtain longer tracks. 

\begin{figure}[htbp]
\begin{center}
\hspace*{-0.5cm} \epsfig{figure=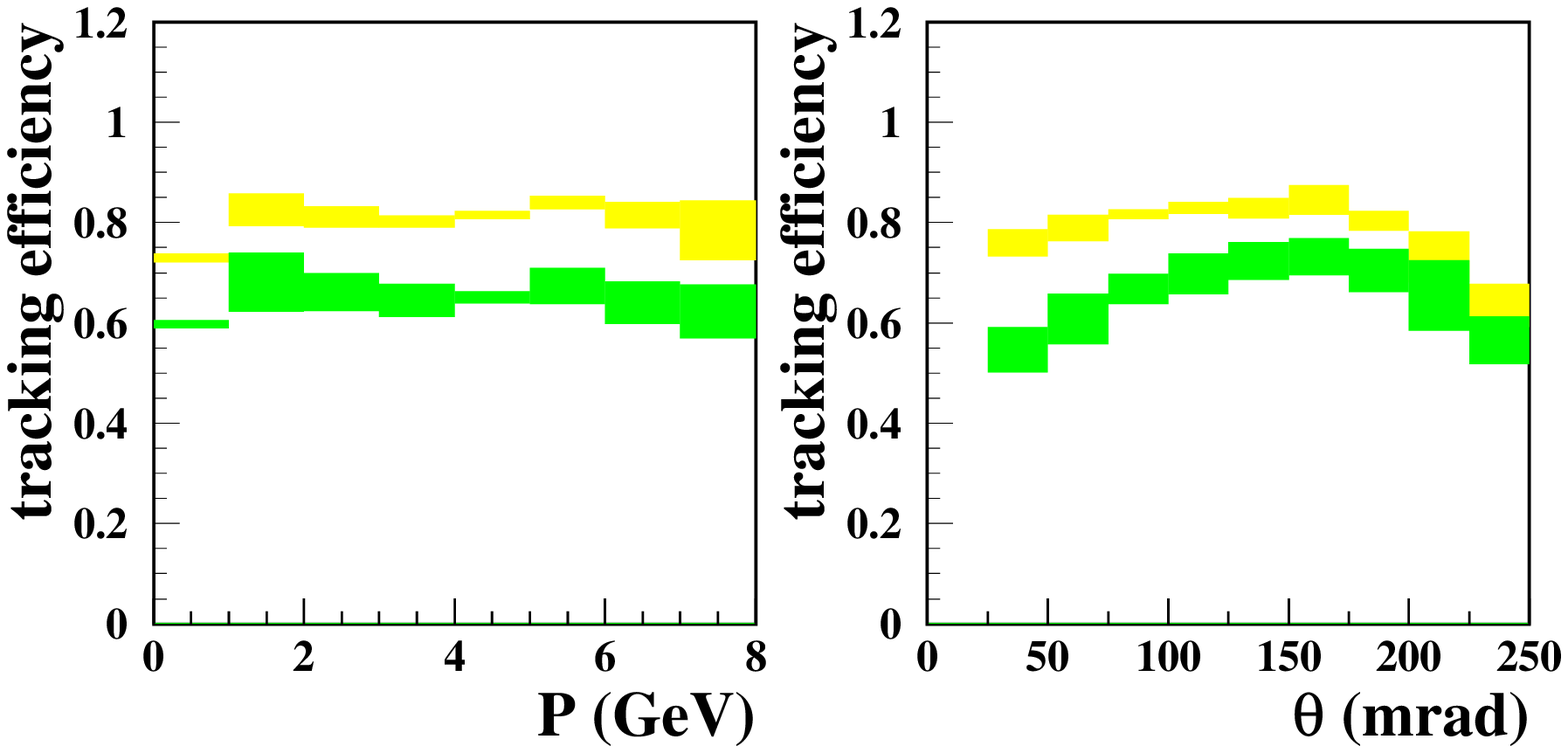,width=11.5cm} 
\epsfig{figure=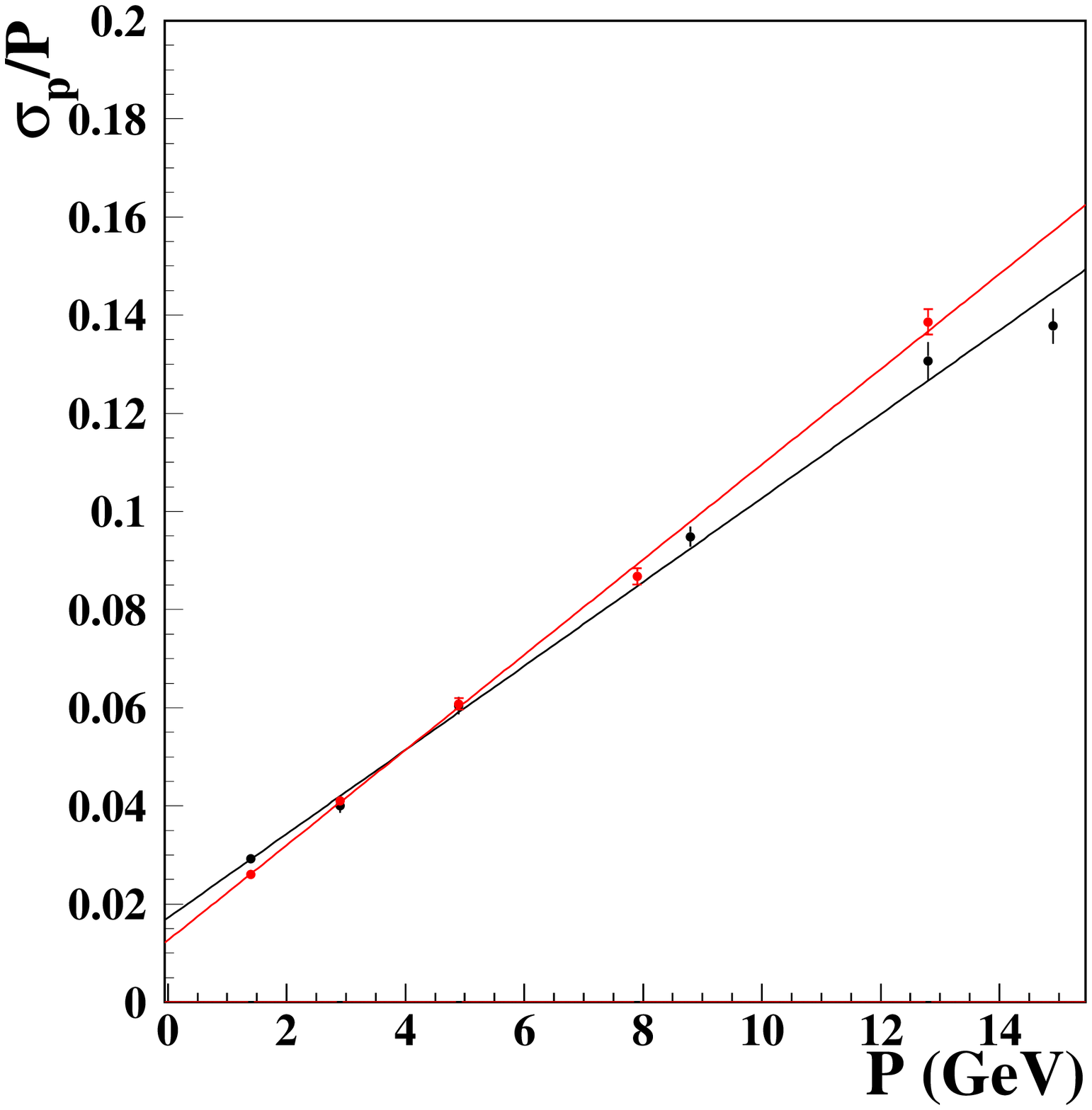,width=5.5cm, height=5cm}
\epsfig{figure=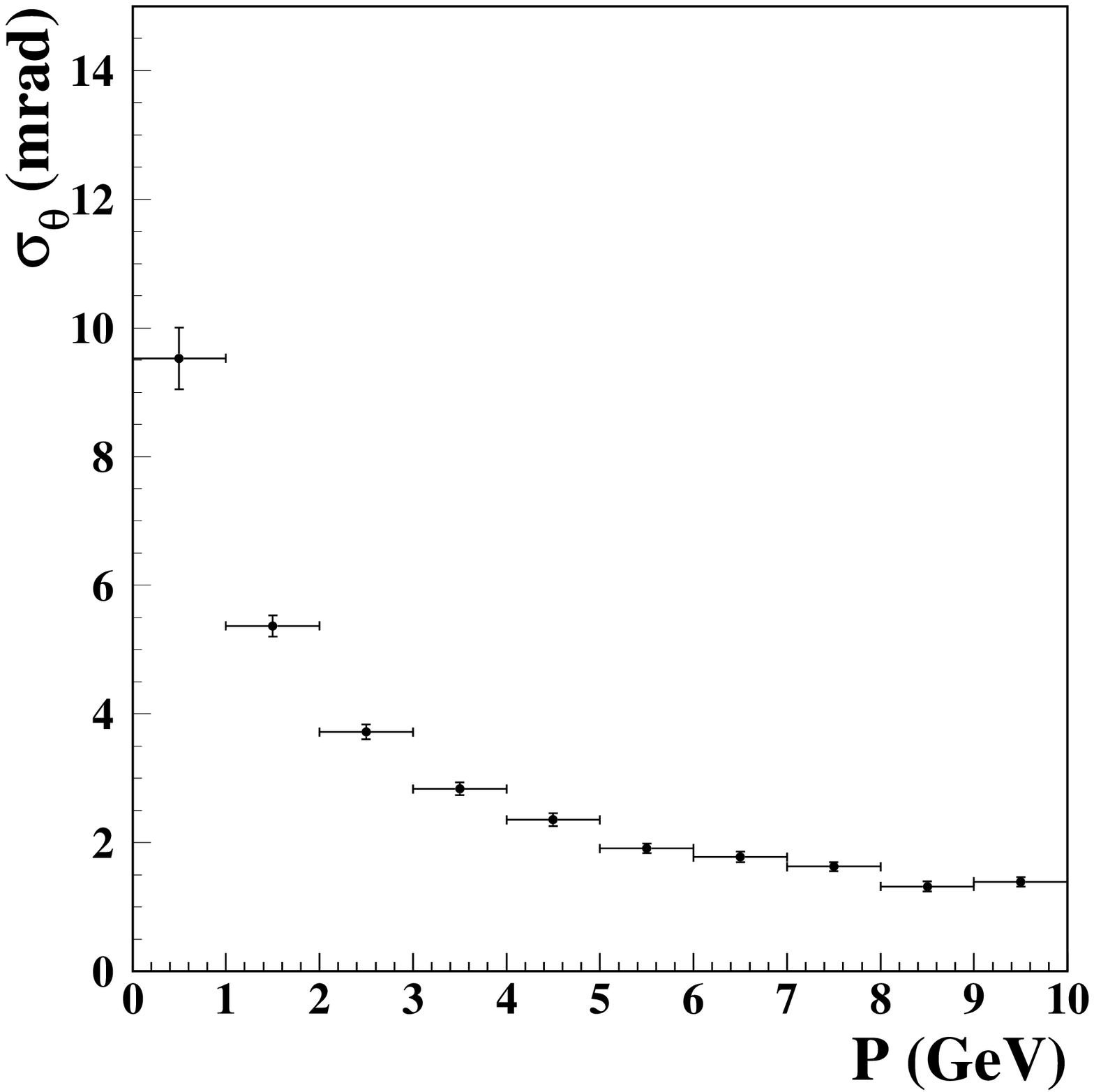,width=5.5cm, height=5cm}
\caption{On the top, tracking efficiency as a function of momentum
          (left) and angle (right). The 
         bands represent the maximum difference between 3 Monte Carlo 
         hadron generators (MC1, MC2, MC3). The 
         lower band corresponds to $3D$-$3D$ up-down matches 
         while the upper band includes also $2D$-$3D$ up-down matches. 
               On the bottom-left, momentum resolution for data (dark) and Monte
               Carlo (light). On the bottom-right, angular
               resolution for Monte Carlo. Both correspond to $3D$-$3D$ matches, with 
               no vertex constraint included.
\label{fig:tracking}}
\end{center}
\end{figure}

The momentum measurement is done associating $3D$ downstream segments with 
$3D$ and $2D$ segments in the upstream module (NDC1).  
This asymmetry is needed to compensate the low tracking efficiency of the upstream chambers, 
due to the presence of a unique module and to the higher 
hit density (proximity to the target), which provokes pattern recognition confusion. 
Fig. \ref{fig:tracking}-bottom shows the momentum and angular
resolutions for $3D$-$3D$ up-down matches. 

The small track separation in NDC1 induces a huge correlation 
between particles that implies a hadron model dependent tracking
efficiency. This is a potential source of systematic error as shown
in Fig. \ref{fig:tracking}-top. The average efficiency is of the order
of $65\%$ when one consideers only $3D$-$3D$ up-down matches. The situation improves
when $2D$-$3D$ matches are included. 
Recent studies with a third type of tracks, built by matching a $3D$
downstream track with the vertex, 
show a considerable efficiency recovery. As these tracks are independent of NDC1, 
the total efficiency ($3D$-$3D$ + $2D$-$3D$ + $vertex$-$3D$) is nearly model independent.

\section{Particle Identification} \label{sec:pid}

Particle identification in HARP's forward region uses the information 
from the time of flight system (TOF), the threshold cerenkov (CKOV) and the electron 
identifier (EID). Pion/proton 
separation is provided by TOF up to $4.5~GeV/c$, and by the CKOV above $3~GeV/c$. 
Electron/pion separation is covered by the CKOV below $3~GeV/c$ and by the EID 
above $2~GeV/c$. Finally the kaon contamination can  be estimated with the  
CKOV above $3~GeV/c$ and with the (TOF) below this energy. 

\subsection{Time of Flight System} \label{subsec:tof}

Particle identification by time of flight in HARP's forward region 
relies on the combination of particle momenta ($p$) and track length ($L$), measured with  
the forward spectrometer, and
the time-of-flight between a start signal ($t_0$) from
the beam time detectors and a stop signal ($t_w$) from the TOF wall, 
placed at about 10 meters downstream from the target. 
The mass of a particle can be computed from these quantities, 
$m^2 = p^2 \cdot [ ((t_w -t_0) \cdot c /L)^2 - 1 ]$.

The current TOF wall and $t_0$ resolutions are 
$150$ and $70~ps$ respectively. However, the
$t_0$ resolution used for 
the analysis shown in this paper was above $200~ps$, leading to a combined TOF resolution 
of about $270~ps$, which is already better that the design value of $300~ps$. 

\begin{figure}[htbp]
\begin{center}
\epsfig{figure=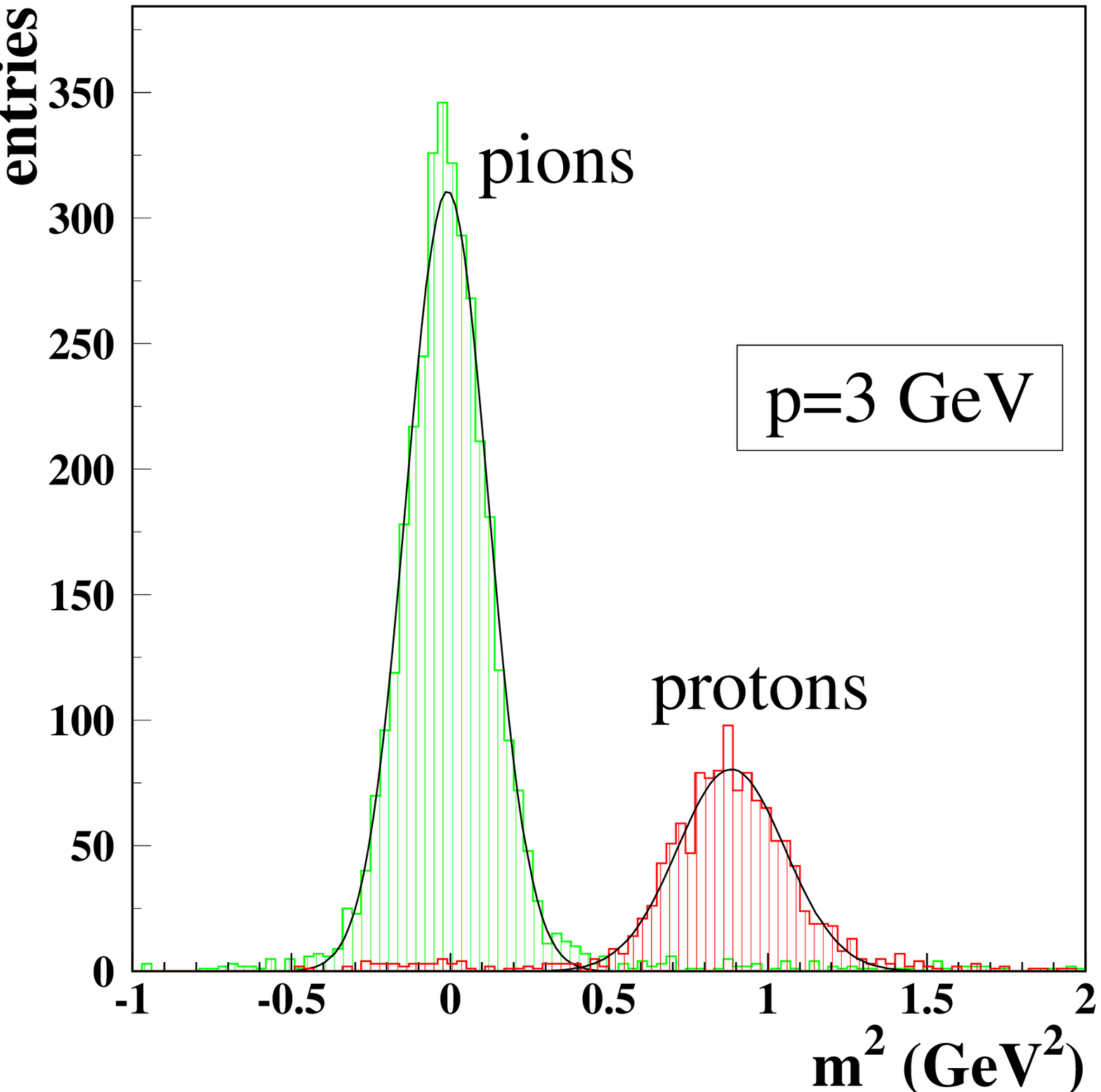,width=5cm, height=5cm}
\epsfig{figure=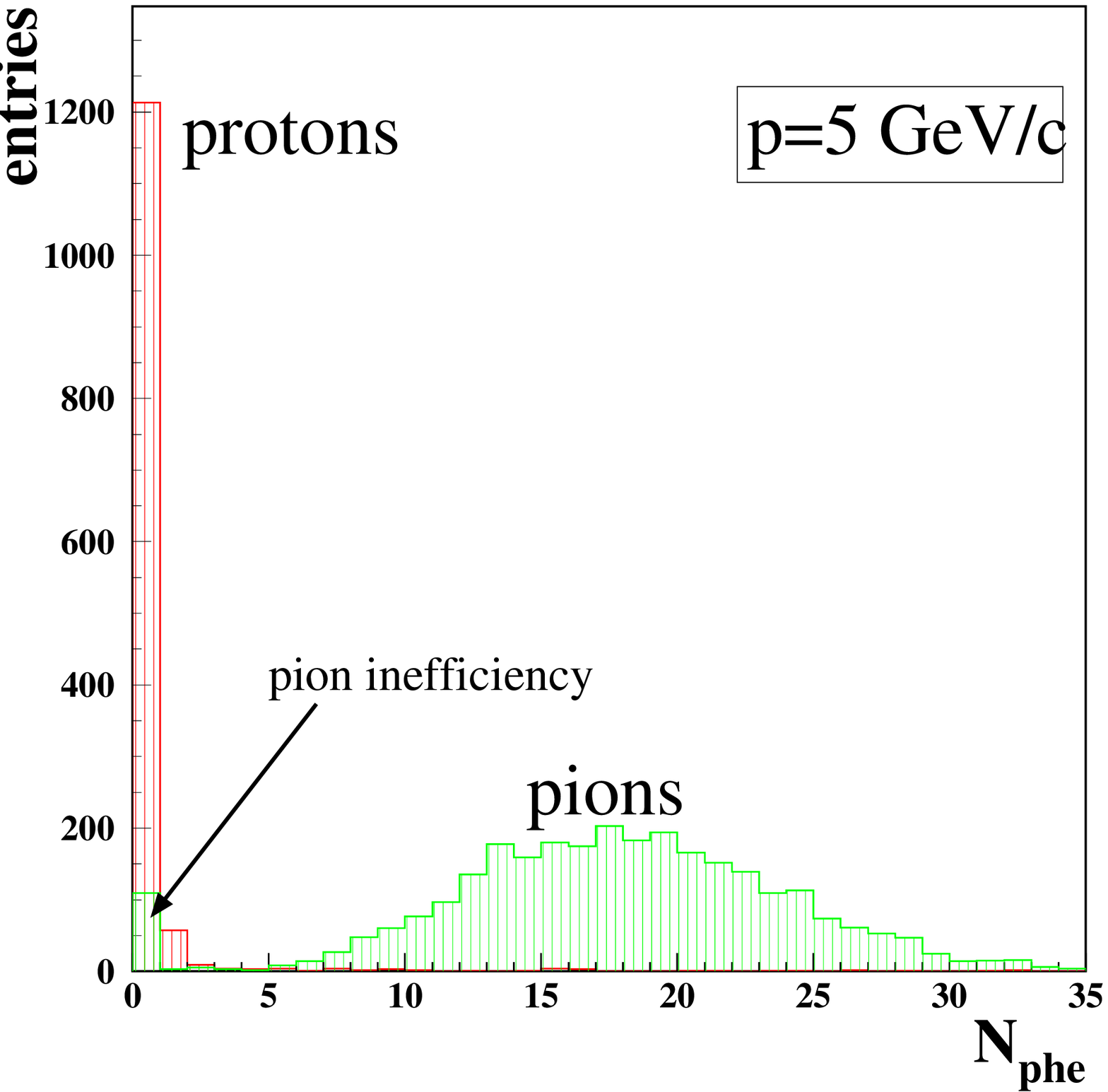,width=5cm, height=5cm}
\epsfig{figure=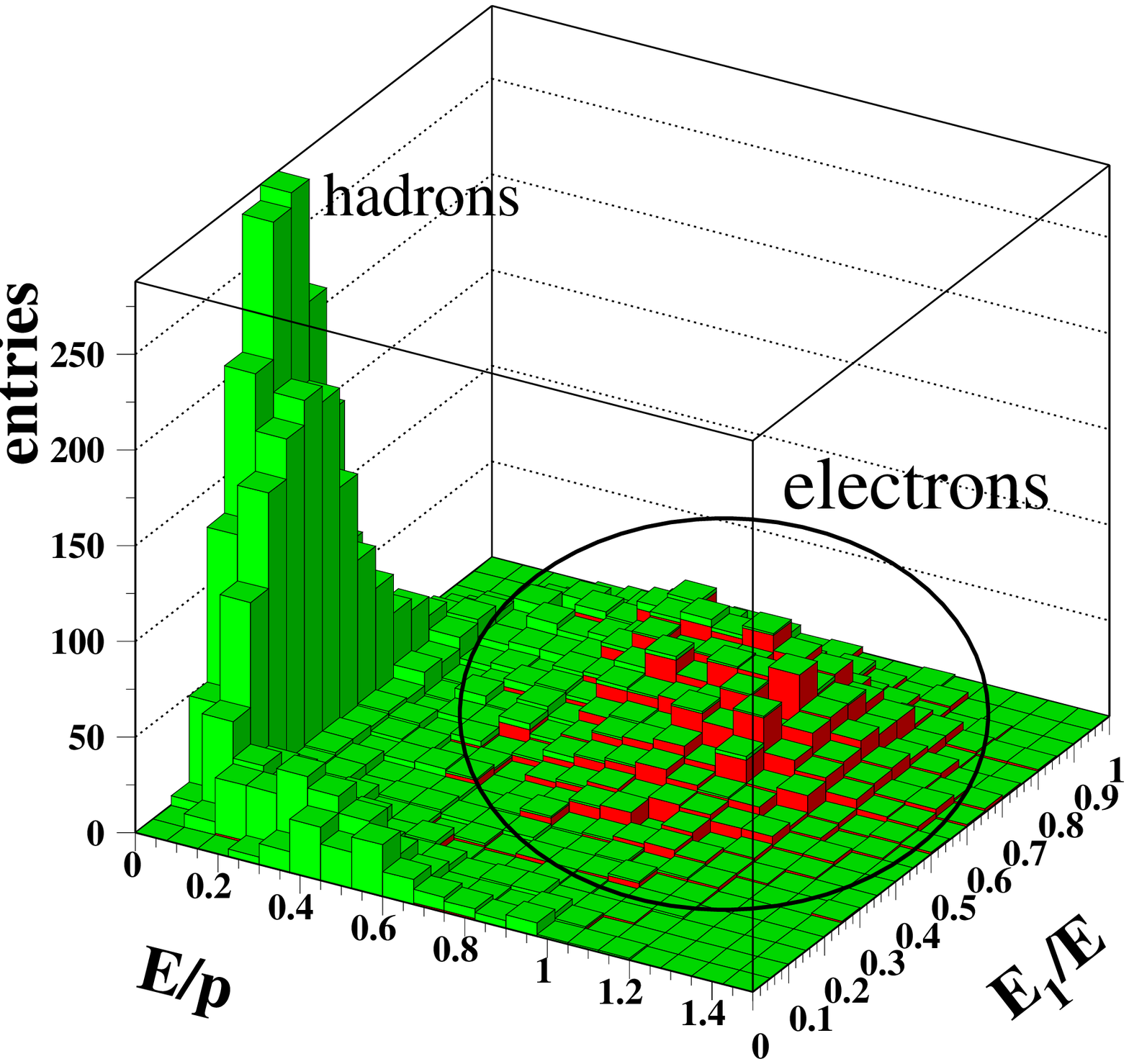,width=5cm, height=5cm}
\caption{On the left, mass squared as measured by the TOF system for $3~GeV/c$ beam particles. 
         On the center, photoelectron yield in the cerenkov produced by beam pions, electrons and protons 
         in a $5~GeV/c$ no target run. 
         On the right, hadron and electron yields as a function of $E_1/E$ and $E/p$ for
         beam particles in a $3~GeV/c$ no target run. $E$ is the sum of the energies deposited 
         in the two calorimeter modules ($E_1+E_2$) and $p$ is the momentum. 
         In all cases pure samples are selected using the beam detectors.  
\label{fig:pid}}
\end{center}
\end{figure}

The TOF is able to provide a $\pi/p$ separation of $\sim5 \sigma$ at
$3~GeV/c$, as can be seen 
in Fig. \ref{fig:pid}-left. In practice, the TOF system will 
provide reasonable $\pi/p$ separation up to $4.5~GeV/c$. The $\pi/k$ capabilities of the 
detector at low energies ($<$$3~GeV/c$) are being studied at the moment
\footnote{It has been recently proposed to use unbiased golden events 
(all beam time detectors firing to gain in $t_0$ resolution) and tracks 
(crossing the overlap region between counters in the TOF wall to
  increase the $t_w$ resolution) 
in order to estimate the $\pi/k$ ratio below $3~GeV/c$. 
The $t_0$ resolution obtained in this case 
is close to $130~ps$. Anyway, the kaon contamination at these energies
is known to be very small.}. 

\subsection{Threshold cerenkov} \label{subsec:ckov}


The cerenkov detector is able to identify electrons below the $\pi$ 
saturation ($\sim 4~GeV/c$). Pions and electrons can be separated from
protons and kaons below the kaon threshold at $\sim 9~GeV/c$,  
and protons can be discriminated above this energy. 
Fig. \ref{fig:pid}-center shows the photoelectron yield 
for beam pions, electrons and protons in a $5~GeV/c$ no target run. 
At $5~GeV/c$ pions are close to the saturation regime and cannot be 
distinguished from electrons, which on the other hand are very rare at these energies. 
However, $\pi/p$ separation is highly pure and efficient at $5~GeV/c$ (and above).

\subsection{Electron identifier} \label{subsec:cal}

The electron identifier is made of two calorimeter planes 
reused from the CHORUS experiment and described elsewhere \cite{chorus}.  
It was designed to provide electron-pion separation
when low energy ($<3~GeV/c$) charged pions, accompanied by knock-on
electrons, are occasionally 
identified as electrons by the cerenkov counter. It also serves to 
identify electrons at high energy, when the cerenkov has lost its pion/electron separation 
capabilities. The detector performance is summarised in Fig.~\ref{fig:pid}-right.

\section{The analysis} \label{sec:analysis}

\begin{figure}[htbp]
\begin{center}
\epsfig{figure=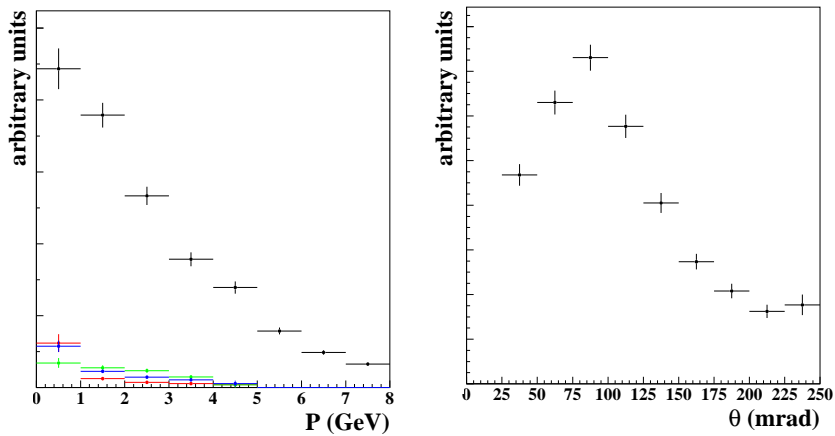,width=12cm}
\caption{Raw pion yield corrected by tracking and pion identification
  efficiencies. On the left, the $p/\pi$ misidentification background
  for the TOF system is also shown (for the three hadron generators).  
  The number of  Monte Carlo events used to compute the tracking
  efficiency was of the order of 50000. This explains the large
  statistical errors in these figures. 
\label{fig:cross}}
\end{center}
\end{figure}

One sixth ( 1 million events ) of the ``K2K thin target'' data has been analysed. 
The unnormalised pion production differential cross section can be computed as follows
\footnote{Bin migration effects are not considered for the moment. Preliminary studies 
show a small effect.}:

\begin{equation}
\sigma_i^\pi = \frac{1}{\varepsilon_i^{acc}} 
               \frac{1}{\varepsilon_i^{track}}
        \frac{1}{\varepsilon_i^{\pi}} 
        \eta_i^{\pi} \cdot
        N_i^{\pi},  
\label{eq:cross}
\end{equation}

\noindent
where the index $i$  corresponds to $(p,\theta)$ bins.  
$\varepsilon_i^{acc}$ is the geometrical acceptance, 
$\varepsilon_i^{track}$ is the tracking efficiency, 
$\varepsilon_i^{\pi}$ is the pion identification efficiency, 
$\eta_i^{\pi}$ is the pion purity and $N_i^{\pi}$ is the
observed pion yield. The pion purity is defined as 
$\eta_i^{\pi} = (N_i^{\pi}-N_i^{bkg}) / N_i^{\pi}$, 
where  $N_i^{bkg}$ is pion misidentification background.
Fig. \ref{fig:cross} shows the raw pion yield corrected
by the tracking and pion identification efficiencies  
($\frac{1}{\varepsilon_i^{track} \cdot\varepsilon_i^{\pi}} N_i^{\pi}$) as a function of 
$p$ and $\theta$. 

The current activities in the context of this forward analysis 
follow the line of recovering tracking efficiency
by using the matching between the downstream tracks and the vertex. 
The most recent results obtained with the improved tracking 
are very encouraging as they are nearly model independent. This will 
allow a considerable reduction of the systematic error.



\section{Conclusions} \label{sec:conclu}

We have described the current performance of the HARP apparatus and
our first physics analysis. This analysis is based on specific data
taken by HARP to improve the neutrino flux calculation for the K2K experiment.
Particle identification and tracking of forward going particles is well 
under control. This has permitted an initial estimation of the
differential raw pion yield and its efficiency correction. 
The analysis is still in progress and will lead to conclusive results shortly.

\section*{Acknowledgments}

I would like to thank the whole HARP Collaboration for their help and support. 
Special thanks to our K2K and MiniBooNE colleagues, Issei Kato, Linda Coney, 
Geoff Mills and Dave Schmitz, who have intensely collaborated to produce these 
results. I'm also grateful to Malcolm Ellis for his invaluable help,
including the English corrections of this article.

\section*{References}

\end{document}